\documentclass[a4paper,12pt]{article}
\usepackage[utf8]{inputenc}
\usepackage[english]{babel}
\usepackage{authblk}
\usepackage{graphicx}
\usepackage{mathptmx}
\usepackage[singlespacing]{setspace}
\usepackage[headheight=1in,margin=1in]{geometry}
\usepackage{fancyhdr}
\usepackage{lipsum}
\usepackage[round]{natbib}
\usepackage{comment}
\usepackage{booktabs}
\usepackage{multirow}
\usepackage{array}
\usepackage{graphicx}

\usepackage{multirow}
\usepackage[table,xcdraw]{xcolor}
\usepackage{graphicx}  
\usepackage{array}     

\makeatletter
\def\@maketitle{%
  \newpage
  \begin{center}%
  \let \footnote \thanks
    {\LARGE \@title \par}%
    \vskip 1.5em%
    {\large
      \lineskip .5em%
      \begin{tabular}[t]{c}%
        \@author
      \end{tabular}\par}%
  \end{center}%
  \par
  \vskip 0.1em}
\makeatother
\graphicspath{{images/}}

\title{Can dialogues with AI systems help humans better discern visual misinformation?}

\author[1]{Anku Rani}
\author[1]{Valdemar Danry}
\author[1]{Andy Lippman}
\author[1]{Pattie Maes}
\affil[1]{MIT Media Lab, Massachusetts Institute of Technology (MIT)

Cambridge, MA

\texttt{ankurani@mit.edu}}

\date{}

\begin{document}

\maketitle
\thispagestyle{fancy}

\begin{center}
\textit{Keywords: Visual Misinformation, Human-AI Interaction, AI Dialogue System, Persuasive Dialogue, Media Literacy}
\newline
\end{center}

\section{Abstract}
The widespread emergence of manipulated news media content poses significant challenges to online information integrity. This study investigates whether dialogues with AI about AI-generated images and associated news statements can increase human discernment abilities and foster short-term learning in detecting misinformation. We conducted a study with 80 participants who engaged in structured dialogues with an AI system about news headline-image pairs, generating 1,310 human-AI dialogue exchanges. Results show that AI interaction significantly boosts participants' accuracy in identifying real versus fake news content from approximately 60\% to 90\% (p$<$0.001). However, these improvements do not persist when participants are presented with new, unseen image-statement pairs without AI assistance, with accuracy returning to baseline levels (~60\%, p=0.88). These findings suggest that while AI systems can effectively change immediate beliefs about specific content through persuasive dialogue, they may not produce lasting improvements that transfer to novel examples, highlighting the need for developing more effective interventions that promote durable learning outcomes.

\section{Related Work} 

The proliferation of altered news media content presents major threats to the reliability of online information ecosystems \citep{farouk2024deepfakes, zhang2020overview}. As digital editing tools become more sophisticated and accessible, along with AI-generated content, the creation and dissemination of manipulated images alongside misleading statements have increased exponentially across social media platforms \citep{giansiracusa2021algorithms}. The combination of visual and textual misinformation is particularly persuasive, as research has shown that humans tend to place heightened trust in visual evidence while often lacking the skills to detect subtle manipulations or critically evaluate accompanying headlines \citep{jagadish2024detection, doi:10.1073/pnas.1806781116, doi:10.1126/science.aap9559}.

\vspace{2mm}

Recent advances in AI have demonstrated promising capabilities in detecting manipulated media content \citep{akram2023empirical, wang-etal-2023-seqxgpt, sadasivan2023can}. However, the role of AI systems extends beyond mere detection. They have the potential to serve as interactive tools that can improve human critical thinking and evaluation skills \citep{walter2024embracing, pedro2019artificial, costello2024durably}. Although existing research has extensively explored automated detection systems \citep{huang-etal-2024-miragenews, wang-etal-2023-seqxgpt, yang-etal-2024-survey}, only recently has attention been paid to how AI can be used as a dialogue partner to improve the human ability to discern truth in the news media. In an experiment with 2,190 conspiracy believers, researchers found that dialogues with an AI chatbot could durably reduce beliefs in conspiracy theories \citep{costello2024durably}. However, it is unclear how these findings generalize to other modalities, such as images with statements, which currently are estimated to make up about 84\% of posts on social media \citep{yang2023visual}. In this research, we study \textbf{whether the diminished beliefs in false information were caused by the persuasiveness of the AI chatbot or its ability to teach users to better discern misinformation in the future.}

\section{Study Design}
To evaluate this research question, we conducted a study with 80 participants interacting with an AI system developed using a prompt similar to previous work on AI dialogues and conspiracy theories \citep{costello2024durably}, and adapted to fake image detection \citep{jia2024can}. We report the accuracy, request rejection rate, and accuracy of non-rejected requests for three prompts using GPT-4o in Table \ref{tab: Prompting} under four conditions: (i) Google search on headline and images passed to GPT-4o for artifact detection, (ii) Google search on headline and no images were passed to GPT-4o, (iii) No Google search on headline and images were passed to GPT-4o, and (iv) No Google search on headline and no images were passed to GPT-4o. Overall accuracy is defined as the model's ability to correctly differentiate between real and fake news headlines and their associated images across the four conditions. Rejection rate is defined as the frequency at which the model declines to provide an answer for a given prompt and non-rejected accuracy is defined as the model's accuracy in the cases where it responded.

\begin{figure*}[t]
    \centering
    \includegraphics[width=1\textwidth]{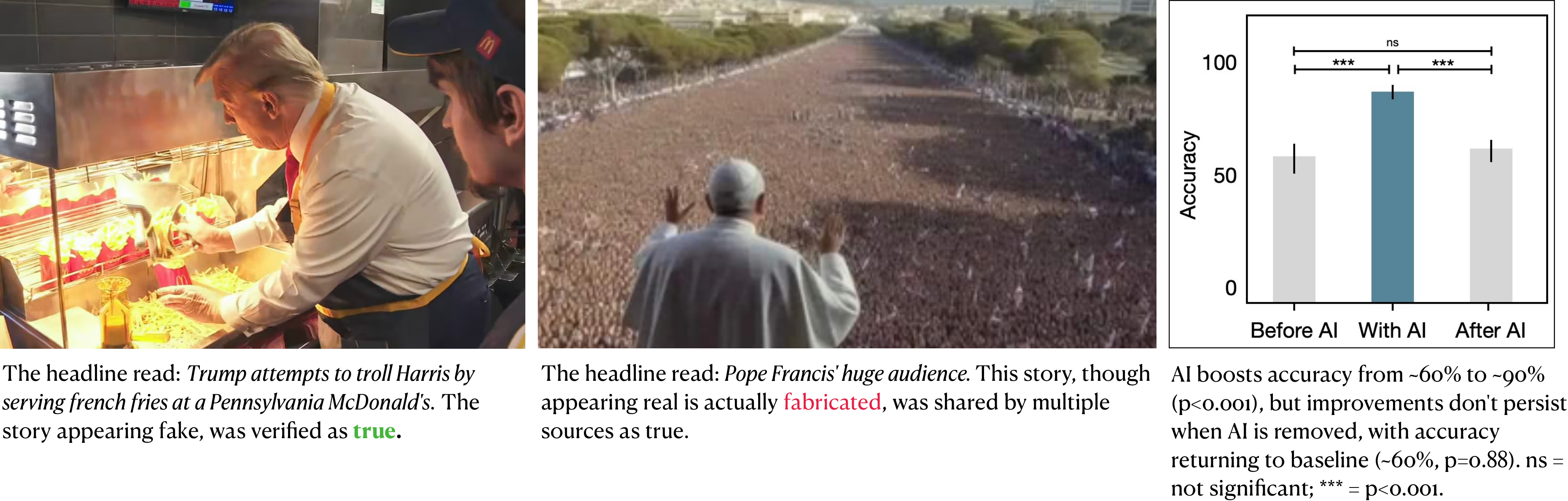}
    \caption{Examples of real and fabricated news stories alongside performance metrics.}
    \label{fig:all}
\end{figure*}

\vspace{2mm}
We compiled a dataset consisting of eight news headline-image pairs that challenged conventional credibility assessment. The dataset included two categories of news content: (1) authentic stories with seemingly implausible elements, and (2) fabricated content crafted to appear credible. This careful selection of materials was designed to test participants' ability to evaluate news content beyond surface-level plausibility judgments. During the intervention phase, participants were presented with four random statement-image pairs from this dataset. One illustrative example features Donald Trump at McDonald's, which is a real image that many participants initially perceived as fabricated.

\begin{figure*}[t]
    \centering
    \includegraphics[width=1.02\textwidth]{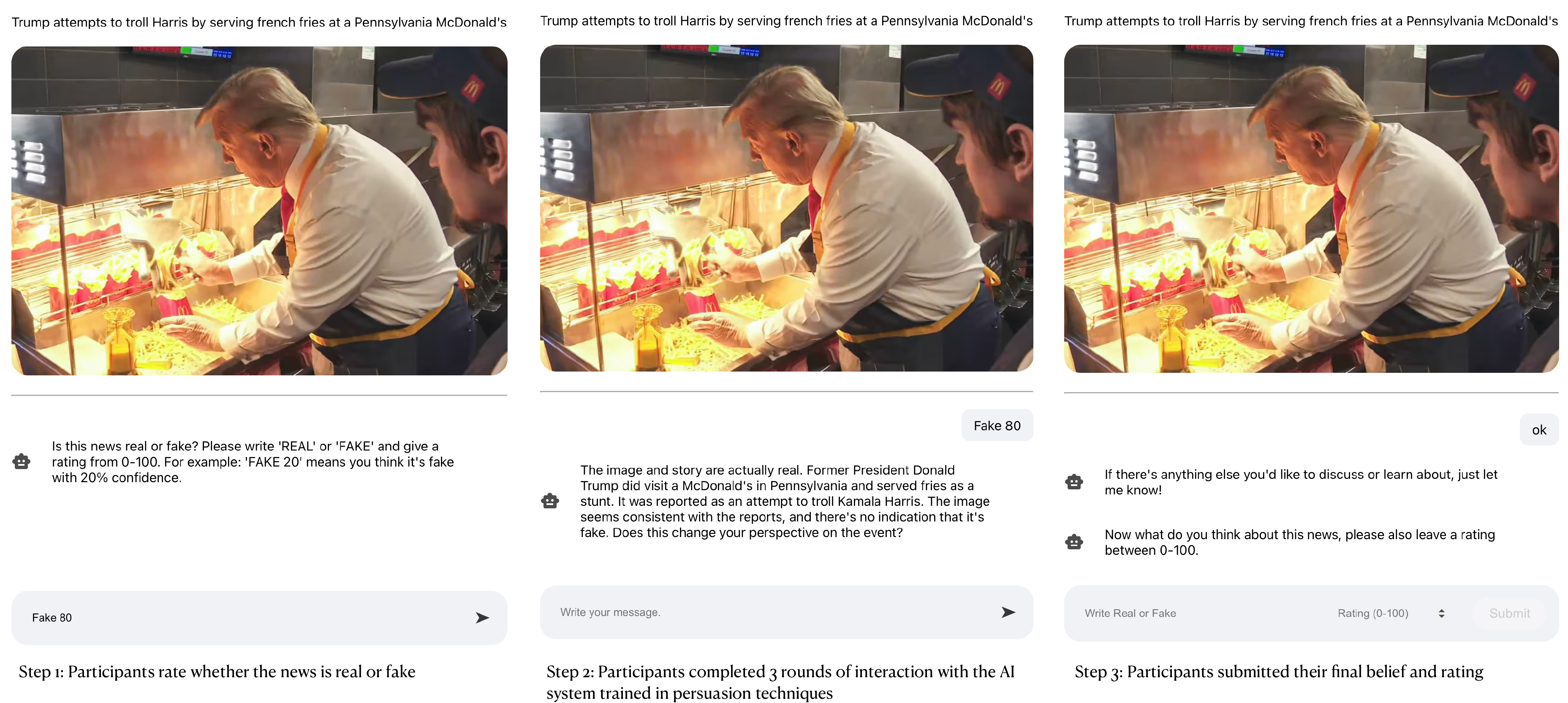}
    \caption{Interaction based on the AI system trained on Persuasion. After participants report whether they have seen the news or not, they interact with the AI system. In step 1: participants rate their beliefs which is treated as initial accuracy. Step 2 shows one interaction and response from the system based on the participant's response. After three rounds of interaction, participants rate their belief which is treated as final accuracy.}
    \label{fig:Interaction}
\end{figure*}

\vspace{2mm}
In the experiment, prior to each AI dialogue on image and statement pairs, participants rated their initial belief in the statement and image pair. Next, each participant engaged in three rounds of structured dialogue with the AI system for each headline-image pair as shown in Figure \ref{fig:Interaction}, culminating in a final authenticity assessment where they rated their belief in the news authenticity. The study generated 1,310 human-AI dialogue exchanges across all participants.

\begin{table}[t]
\setlength{\tabcolsep}{4pt}  
\resizebox{\textwidth}{!}{%
\begin{tabular}{p{6.5cm}lccccc}
\hline
Prompt & Metric & \begin{tabular}[c]{@{}c@{}}Condition 1\\ google search on headline\\ + image on GPT-4o\end{tabular} & \begin{tabular}[c]{@{}c@{}}Condition 2\\ google search\\ + no image\end{tabular} & \begin{tabular}[c]{@{}c@{}}Condition 3\\ no google search\\ + image\end{tabular} & \begin{tabular}[c]{@{}c@{}}Condition 4\\ no google search\\ + no image\end{tabular} \\ \hline

\multirow{3}{*}{\begin{tabular}[t]{@{}l@{}}Prompt 1: News forensic expert\\(Artifact detection only)\end{tabular}} 
& Overall Accuracy & 71.43\% & 78.57\% & 35.71\% & 0.00\% \\
& Rejection Rate & 21.43\% & 21.43\% & 50.00\% & 92.86\% \\
& Accuracy (Non Rejected) & 90.91\% & 100.00\% & 71.43\% & 0.00\% \\ \hline

\multirow{3}{*}{\begin{tabular}[t]{@{}l@{}}Prompt 2: Persuasion\\(Belief change only)\end{tabular}}
& Overall Accuracy & 71.40\% & 78.57\% & 28.57\% & 0.00\% \\
& Rejection Rate & 14.20\% & 14.20\% & 78.57\% & 92.00\% \\
& Accuracy (Non Rejected) & 83.30\% & 91.60\% & 66.66\% & 0.00\% \\ \hline

\multirow{3}{*}{\begin{tabular}[t]{@{}l@{}}Prompt 3: Combined\\(News Forensic Expert + Persuasion)\end{tabular}}
& Overall Accuracy & \cellcolor[HTML]{90EE90}100\% & 78.5\% & 42.8\% & 0\% \\
& Rejection Rate & \cellcolor[HTML]{90EE90}0\% & 14.20\% & 21.42\% & 71.4\% \\
& Accuracy (Non Rejected) & \cellcolor[HTML]{90EE90}100\% & 91.60\% & 54.45\% & 0\% \\ \hline

\end{tabular}
}
\caption{Comparison of Different Prompt Conditions and Their Impact on Accuracy and Rejection Rates. These prompts are described in Figure \ref{fig:Prompts}.}
\label{tab: Prompting}

\end{table}

\section{Results}
To assess learning during the AI system interaction, we also measured participants' final accuracy performance on four unseen image and statement pairs without AI assistance after their interactions with the chatbot. The accuracy results are reported in Figure \ref{fig:all}.

To analyze differences in accuracy measures, we conducted a one-way analysis of variance (ANOVA) which revealed significant differences among initial accuracy, final accuracy, and learning accuracy ($F = 39.28$, $p < 0.001$). Subsequent pairwise post-hoc t-tests showed that final accuracy was significantly higher than initial accuracy ($t = -7.16$, $p < 0.001$), indicating participants improved their performance over time. Learning accuracy was not significantly different from initial accuracy ($t = -0.15$, $p = 0.88$), but was significantly lower than final accuracy ($t = 10.53$, $p < 0.001$). These results suggest that while participants achieved better final performance on pairs in which they had engaged in dialogues compared to their initial accuracy on the same pairs, their learning accuracy was not significantly different than their initial performance levels.

\section{Future Work}
Our future work will focus on overcoming the limitation of the current experiment where participants had limited interaction with the AI. We believe having prolonged exposure to the intervention system might affect learning. We also measured learning effects immediately after the intervention and would like to measure them again after one week and one month. We will also explore a second hypothesis: that AI systems optimized for critical thinking strategies not only decrease belief in misleading content but also produce sustained improvements in human evaluation skills. This expanded analysis will examine whether incorporating critical thinking frameworks into AI dialogue systems leads to both immediate reductions in fake news susceptibility and enhanced learning outcomes that transfer to novel examples.


\bibliography{references}

\newpage

\begin{figure*}[t]
    \centering
    \includegraphics[width=1\textwidth]{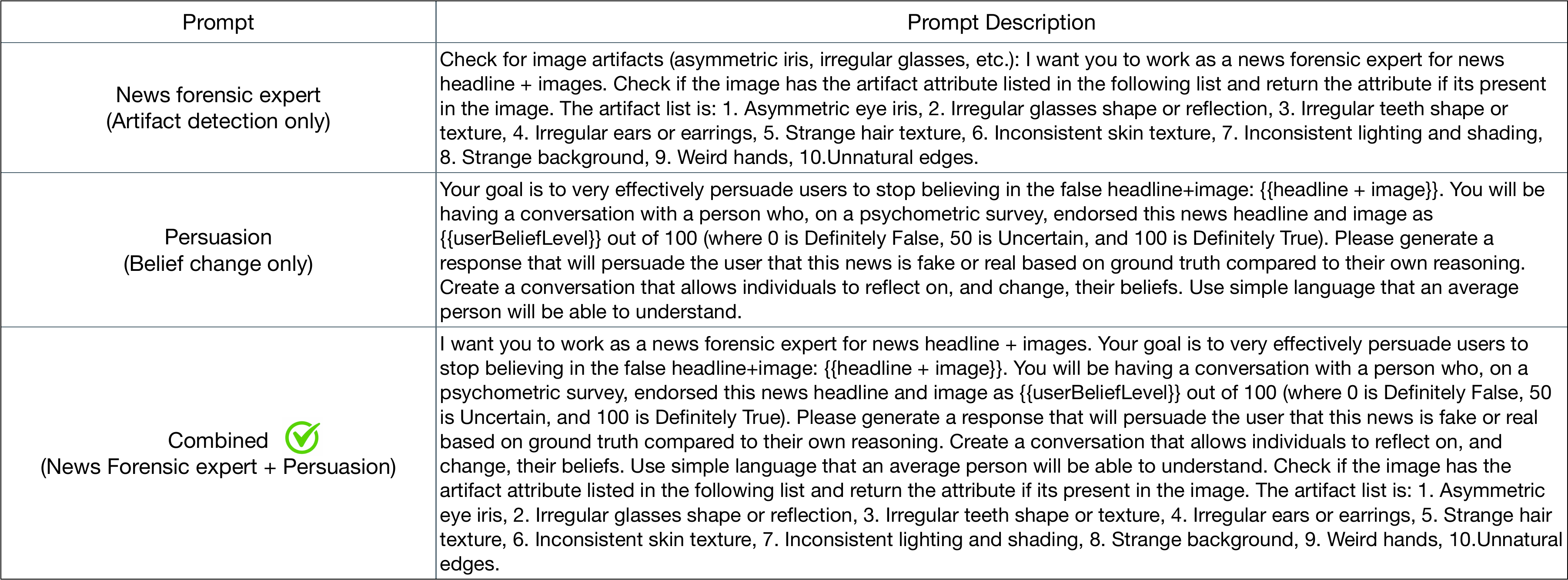}
    \caption{Prompting Strategies Description for News headline and image Credibility Assessment. The optimal strategy (Prompt 3) integrates approaches for artifact detection and persuasion to enhance human-AI dialogue to distinguish between real and fake news headline image pairs.}
    \label{fig:Prompts}
\end{figure*}

\end{document}